\documentclass[prd,amsmath, twocolumn,floatfix,amssymb, preprintnumbers, nofootinbib, superscriptaddress]{revtex4} 

\usepackage{hyperref}
\usepackage{epsfig,dcolumn}
\usepackage{graphicx}
\usepackage{comment} 
\usepackage{verbatim}
\usepackage{color}
\usepackage{graphicx}
\usepackage{bm}
\usepackage{ifpdf}
\usepackage{multirow}
\usepackage{soul} 

\def\lsim{\mathrel{\rlap{\lower4pt\hbox{\hskip1pt$\sim$}}
    \raise1pt\hbox{$<$}}}
\def\gsim{\mathrel{\rlap{\lower4pt\hbox{\hskip1pt$\sim$}}
    \raise1pt\hbox{$>$}}}
    
\begin{document}

\title{Theoretical analysis of the $\gamma\gamma \to \pi^0 \eta$ process}

\author{Igor Danilkin}\email{danilkin@uni-mainz.de}
\affiliation{Institut f\"ur Kernphysik \& PRISMA  Cluster of Excellence, Johannes Gutenberg Universit\"at,  D-55099 Mainz, Germany}
\author{Oleksandra Deineka}
\affiliation{Institut f\"ur Kernphysik \& PRISMA  Cluster of Excellence, Johannes Gutenberg Universit\"at,  D-55099 Mainz, Germany}
\affiliation{Department of Physics, Taras Shevchenko National University of Kyiv, 6 Academician Glushkov Avenue., Kyiv 03680, Ukraine}
\author{Marc Vanderhaeghen}
\affiliation{Institut f\"ur Kernphysik \& PRISMA  Cluster of Excellence, Johannes Gutenberg Universit\"at,  D-55099 Mainz, Germany}

\date{\today}

\begin{abstract}
We present a theoretical study of the $\gamma\gamma \to \pi\eta$ process from the threshold up to 1.4 GeV in the $\pi\eta$ invariant mass. For the s-wave $a_0(980)$ resonance state we adopt a dispersive formalism using a coupled-channel Omn\`es representation, while the d-wave $a_2(1320)$ state is described as a Breit-Wigner resonance. An analytic continuation to the $a_0(980)$ pole position allows us to extract its two-photon decay width as $\Gamma_{a_0\to\gamma\gamma}=0.27(4)$ keV.
\end{abstract}

\maketitle

\section{Introduction}

Recently, the Belle Collaboration measured the exclusive hadronic $\pi^0\eta$ production in two-photon collisions \cite{Uehara:2009cf}. The statistics of these new data is more than two orders of magnitude higher than any previous measurements in this channel \cite{Oest:1990ki, Antreasyan:1985wx} and therefore provides valuable information on the nature of scalar $a_0(980)$ and tensor $a_2(1320)$ resonances. In particular it sheds light on the two-photon strength of the $0^{++}$ isovector channel which serves as an important constraint in the light-by-light scattering \cite{Pascalutsa:2012pr} and consequently to the hadronic contribution to the anomalous magnetic moment of the muon $a_\mu = (g - 2)_\mu/2$ \cite{Jegerlehner:2017gek, Benayoun:2014tra, Prades:2009tw, Danilkin:2016hnh}.

The method we use is based on the fundamental principles of the $S$-matrix, i.e. analyticity and unitarity. In this way, final state interactions are fully accounted for. Secondly, there are no unknown parameters.  All the couplings which enter the dispersion integral are fixed from the radiative decays of the vector mesons into pseudoscalar mesons. In this sense, our analysis is different from an earlier work which has a significant amount of unknown parameters and therefore a limited predictive power \cite{Danilkin:2012ua}. In addition to that, the analyticity constraint, which is hard to implement, is frequently discarded in the literature \cite{Oller:1997yg, Achasov:2010kk, Lee:1998mz}.

In the dispersion formalism, there are always contributions from the right- and left-hand cuts \cite{Morgan:1987gv}. While the right-hand cuts of the scattering amplitude are fixed from unitarity, the left-hand cuts lie in the unphysical region and can be approximated by vector-meson exchanges \cite{Danilkin:2012ua, GarciaMartin:2010cw, Moussallam:2013una}. In order to benchmark the proposed treatment for the left-hand cuts, we study the double radiative decay, $\eta \to \pi^0 \gamma \gamma$, which is related to the scattering process by the crossing transformation.

In the description of the scattering process, it is well known that the $a_0(980)$ resonance has a strong coupling to the $K\bar{K}$ channel. Therefore, the coupled-channel dispersion integral was used to implement such rescattering effects through two intermediate kaons \cite{GarciaMartin:2010cw}. In order to determine the pole position and the two-photon coupling of the $a_0(980)$ resonance, the amplitude is analytically continued to the unphysical Riemann sheets. This is particularly important since there is an interplay between elastic and inelastic channels and the structure of that resonance is significantly different from a typical Breit-Wigner form. In contrast, the tensor $a_2(1320)$ resonance is described as a Breit-Wigner resonance, using its experimentally measured two-photon decay width \cite{PDG-2016}.

The paper is organized as follows. In the next section, we summarize the kinematics and discuss the main features of the dispersive framework for the $\gamma\gamma \to \pi \eta$ process. The hadronic input and the role of the left-hand cuts are discussed in Sections~\ref{sec:Hadronic_input} and \ref{sec:Left-hand_cuts}. In Section~\ref{sec:a2} we present the details of the tensor $a_2(1320)$ resonance. The numerical analysis of the $\eta\to \pi^0 \gamma\gamma$ decay is presented in Section~\ref{sec:Decay}. Subsequently, we show our numerical results in Section \ref{sec:Results}. A summary and outlook are presented in Section~\ref{sec:Conclusions}. 

\section{Formalism}
\label{sec:Formalism}
\subsection{Kinematics and partial wave expansion}
\label{sec:Kinematics}
The photon fusion reaction $\gamma\gamma \to \pi \eta$ is described by the $T$-matrix element, which is related to the $S$-matrix element as $S =1 + i T$, and which can be written as
\begin{eqnarray}
&&\langle \pi(p_1)\eta(p_2)| T| \gamma(q_1,\lambda_1)\gamma(q_2,\lambda_2)\rangle\nonumber \\
 &&= (2\pi)^4\,\delta^{(4)}(p_1+p_2-q_1-q_2)\,H_{\lambda_1 \lambda_2}\,,
\end{eqnarray}
where $\lambda_{1,2}=\pm 1$ are the photon helicities. The particle momenta $q_{1,2}$ and $p_{1,2}$ are related to the Mandelstam variables by $s=(q_1+q_2)^2$, $t=(p_1-q_1)^2$ and $u=(p_1-q_2)^2$ which satisfy the relation $s+t+u=m_{\pi}^2+m_{\eta}^2$. The helicity amplitudes can be expressed in terms of the complete set of invariant amplitudes $F_{1,2}(s,t)$,
\begin{eqnarray}\label{HelicityAmplitudes}
H_{\lambda_1 \lambda_2}= \epsilon_\mu(q_1,\lambda_1)\,\epsilon_\nu(q_2,\lambda_2)\,\left[F_1(s,t)L^{\mu\nu}_1+F_2(s,t)L^{\mu\nu}_2\right],
\nonumber
\end{eqnarray}
where $\epsilon_\nu(q_{1,2},\lambda_{1,2})$ are the polarization vectors of the initial photons. The main constraint on the Lorentz tensors in Eq.(\ref{HelicityAmplitudes}) is that the invariant amplitudes should be free from kinematic singularities \cite{Bardeen:1969aw} and therefore should satisfy Mandelstam analyticity \cite{Mandelstam:1958xc,Mandelstam:1959bc}. We note, however, that the choice of a particular set of tensors $L_{1,2}^{\mu\nu}$ is not unambiguous\footnote{One can always introduce a new set of Lorentz tensors $\tilde{L}_{1,2}^{\mu\nu}$ as a linear combination of the given basis tensors $L_{1,2}^{\mu\nu}$ without spoiling kinematic and gauge invariance constraints.}. We use the decomposition from \cite{Danilkin:2012ua}
\begin{eqnarray}\label{Eq:Lorez_structures}
L_1^{\mu\nu}&=&q_1^{\nu}\,q_2^{\mu}-(q_1\cdot q_2)\,g^{\mu\nu}\,,\\
L_2^{\mu\nu}&=&(\Delta^2\,(q_1\cdot q_2)-2\,(q_1\cdot
\Delta)\,(q_2\cdot
\Delta))\,g^{\mu\nu}\nonumber \\
&&-\Delta^2\,q_1^\nu\,q_2^\mu
    -2(q_1\cdot q_2)\,\Delta^{\mu}\,\Delta^{\nu}\nonumber\\
    &&+2(q_2\cdot \Delta)\,q_1^{\nu}\,\Delta^{\mu}+2(q_1\cdot \Delta)\,q_2^{\mu}\,\Delta^{\nu}\,,\nonumber
\end{eqnarray}
with $\Delta=p_1-p_2$. These relations satisfy the Ward identities $q_{1\mu}L^{\mu\nu}_{1,2}=0$, $q_{2\nu}L^{\mu\nu}_{1,2}=0$, and also have the orthogonality property $L_{1}^{\mu\nu}\,L_{2,\mu\nu}=0$ which proves to be convenient for further calculations. 

From the helicity amplitudes, it is straightforward to obtain the differential cross section
\begin{equation}\label{Eq:Cross_section}
    \frac{d\sigma}{d\cos\theta}=\frac{\beta_{\pi\eta}(s)}{64\,\pi\,s}\,\left(|H_{++}|^2+\,|H_{+-}|^2\right)\,,
\end{equation}
where
\begin{equation}
\beta_{ij}(s)=\frac{1}{s}\sqrt{[s-(m_i+m_j)^2][s-(m_i-m_j)^2]}\,.
\end{equation}

When studying low-lying resonances it is useful to perform a partial wave (p.w.) expansion of the helicity amplitudes with fixed isospin $(I)$ \cite{Jacob:1959at}:
\begin{equation}
H_{I,\lambda_1\lambda_2}(s,t)=\sum_{\textrm{even }J\geq0}(2J+1)\,h^{J}_{I,\lambda_1,\lambda_2}(s)\,d_{\lambda_1-\lambda_2,0}^{J}(\theta)\,,\nonumber \\
\end{equation}
where $d_{\lambda,\bar{\lambda}}^{J}(\theta)$ are Wigner rotation functions and $\theta$ is the center-of-mass scattering angle in the $x−z$ reaction plane, where we choose the z-axis along the photon directions. Note that the same p.w. decomposition holds for $\gamma\gamma \to K\bar{K}$ helicity amplitude $K_{I,\lambda_1,\lambda_2}$, denoting the $\gamma\gamma \to K\bar{K}$ p.w. amplitudes as $k^{J}_{I,\lambda_1,\lambda_2}$ in the following. The isospin transformations for the $\gamma\gamma \to K\bar{K}$ are
\begin{eqnarray}\label{IsospinK}
K_{0,\lambda_1,\lambda_2}=-\frac{1}{\sqrt{2}}(K^c_{\lambda_1,\lambda_2}+K^n_{\lambda_1,\lambda_2})\,,\\
\nonumber
K_{1,\lambda_1,\lambda_2}=-\frac{1}{\sqrt{2}}(K^c_{\lambda_1,\lambda_2}-K^n_{\lambda_1,\lambda_2})\,,
\end{eqnarray}
where $K^c$ and $K^n$ correspond to the charged and neutral amplitudes, respectively. The $\gamma\gamma \to \pi^0\eta$ process, in turn, is a pure $I=1$ process.

\subsection{Coupled-channel Omn\`es representation}
\label{sec:Omnes_representation}
It is well known that the coupled-channel final state interaction in the s-wave isovector sector is very strong and necessary in order to properly describe the $a_0(980)$ resonance.  Assuming Mandelstam analyticity, the p.w. amplitudes $h^J_{\lambda_1,\lambda_2}(s)$ should satisfy p.w. dispersion relations. We follow the formalism outlined in Ref. \cite{GarciaMartin:2010cw} for the case of $\gamma \gamma \to \pi\pi, K\bar{K}$ scattering where the $K\bar{K}$ channel is needed for a proper description of the $f_0(980)$ resonance. In \cite{GarciaMartin:2010cw} the dispersion relation is written for the function $\Omega^{-1}(s)(h(s)-h^{Born}(s))$, which contains both left- and right-hand cuts. The particular form splits the well-known Born left-hand cut ($s<0$) from other heavier intermediate $t$- and $u$- channel state contributions ($s<s_L$). For the $I=1$, s-wave scattering we write a once-subtracted dispersion relation
\begin{eqnarray}\label{h_CC}
&&\left(\begin{array}{c}h^{0}_{1,++}\\
k^{0}_{1,++}\end{array}\right)=\left(\begin{array}{c}0\\
k^{0, Born}_{1,++}(s)\end{array}\right) +\Omega_{1}^{0}(s) \left[\left(\begin{array}{c}a\\b\end{array}\right)\right. 
\\&&+\left.\frac{s-s_{th}}{\pi}\int_{-\infty}^{s_L}
\frac{ds'}{s'-s_{th}}\,\frac{\Omega_{1}^{0}(s')^{-1}}{s'-s}\left(\begin{array}{c}\text{Disc}\,h^0_{1,++}(s')\\\text{Disc}\,\bar{k}^0_{1,++}(s')\end{array}\right)\right. \nonumber \\
 &&-\left.\frac{s-s_{th}}{\pi}\int_{s_{th}}^{\infty}\frac{ds'}{s'-s_{th}}\,\frac{\text{Disc}\,\Omega_{1}^{0}(s')^{-1}}{s'-s}\left(\begin{array}{c}0\\ k^{0,Born}_{1,++}(s')\end{array}\right)
\right]\nonumber
\end{eqnarray}
where $s_{th}=(m_\pi+m_\eta)^2$ and $\bar{k}(s)$ is a non-Born part of $k(s)$. The hadronic Omn\`es matrix
\begin{equation}\label{Omnes_CC}
\Omega_{1}^{0}(s)=\left(
\begin{array}{cc}
\Omega_{1}^{0}(s)_{\pi\eta \to\pi\eta} & \Omega_{1}^{0}(s)_{\pi\eta \to K\bar{K}}\\
\Omega_{1}^{0}(s)_{K\bar{K} \to \pi\eta} & \Omega_{1}^{0}(s)_{K\bar{K} \to K\bar{K}}
\end{array}
\right)
\end{equation}
normalized as $\Omega(s_{th})=1$ and satisfies the following unitarity condition
\begin{eqnarray}\label{Omnes_CC_Unitarity}
\text{Disc}\,\Omega_{1}^{0}(s)&=&\frac{1}{2i}\left(\Omega_{1}^{0}(s+i\epsilon)-\Omega_{1}^{0}(s-i\epsilon)\right)\nonumber \\&=&t_{1}^{0}(s)\,\rho(s)\,\Omega_{1}^{0*}(s),\quad s>s_{th}\,.
\end{eqnarray}
Here $t_{1}^{0}(s)$ is the hadronic p.w. scattering matrix and
\begin{equation}
\rho(s)=\frac{1}{16\pi} \left(
\begin{array}{cc}
\beta_{\pi\eta}(s)\theta(s-s_{th}) & 0\\
0 & \beta_{K\bar{K}}(s)\theta(s-4m_K^2)
\end{array}
\right)
\end{equation}
is the phase space matrix.

\subsection{Hadronic input}
\label{sec:Hadronic_input}
\begin{figure}[t]
\includegraphics[width =0.5\textwidth]{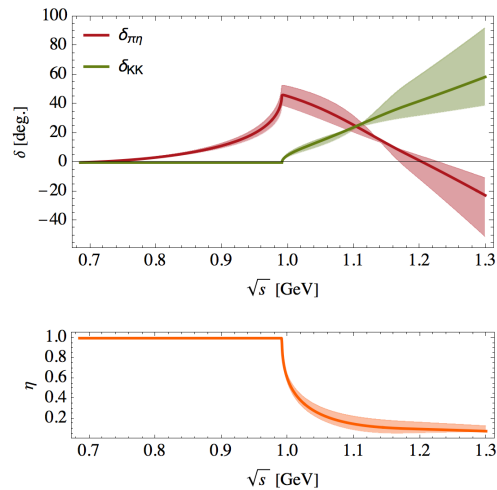}
\caption{$s$-wave phases shifts and inelasticity used in this work, for $\{\pi\eta,K\bar{K}\}$ coupled-channel scattering with $I = 1$. The shaded bands indicate the theoretical uncertainty as discussed in the text.}
\label{fig:phases}
\end{figure}

The photon-fusion reactions are sensitive to the hadronic final state interactions. Therefore, the important input is a proper description of the $\pi \eta$ rescattering processes. In contrast to the $\pi\pi$ scattering, there is no $\pi\eta$ scattering data available, and it is impossible to build a data-driven dispersive solution for the Omn\`es function. However, as it was shown in \cite{Danilkin:2011fz, Danilkin:2012ont} one can apply the recently proposed dispersive summation scheme \cite{Gasparyan:2010xz, Gasparyan:2012km, Danilkin:2010xd} which implements constraints from analyticity and unitarity and is consistent with chiral perturbation theory ($\chi$PT) at low energies. The method is based on the $N/D$ ansatz \cite{Chew:1960iv}, where the set of coupled-channel integral equations for the $N$-function was solved numerically 
\begin{eqnarray}\nonumber
N(s)=U(s)+\frac{s-s_{th}}{\pi}\int_{s_{th}}^{\infty}ds'\frac{N(s')\rho(s')(U(s')-U(s))}{(s'-s_{th})(s'-s)}
\end{eqnarray}
with the input from the suitably constructed conformal mapping expansion 
\begin{eqnarray}
U(s)=\sum_k C_k\,\xi(s)^k\,,
\end{eqnarray}
which parametrizes all contributions coming from the left-hand cuts. The coefficients $C_k$ of this expansion were matched at threshold to the tree level $\chi$PT supplemented with the light vector meson fields. After solving the linear integral equation for $N(s)$, the $D(s)$-function was computed, which is the inverse of the Omn\`es function,
\begin{eqnarray}\nonumber
\Omega^{-1}(s)=1-\frac{s-s_{th}}{\pi}\int_{s_{th}}^{\infty}\frac{ds'}{s'-s_{s_{th}}}\frac{N(s')\rho(s')}{s'-s}\,.
\end{eqnarray}
The final hadronic scattering amplitude was reconstructed by $t(s)=\Omega(s)N(s)$. In Refs. \cite{Danilkin:2011fz, Danilkin:2012ap} it has been shown that one can achieve a reasonable agreement with the existing experimental data of $\pi\pi$ and $\pi K$ scattering and at the same time predict the, yet to be measured, $\pi \eta$ and $K\bar{K}$ scattering. The latter result was not included in \cite{Danilkin:2011fz} and since it is essential for the $\gamma\gamma \to \pi \eta$ reaction we show the $\delta_{\pi\eta}$ and $\delta_{K\bar{K}}$ phase shifts and inelasticity, used in this work, in Fig. \ref{fig:phases}.

We recall that in the approach presented in \cite{Danilkin:2011fz} there are only a few relevant and known parameters. These are the pion decay constant in the chiral limit, the coupling constant of the vector meson into two pseudoscalar mesons (e.g. $\rho\to\pi\pi$) and the parameter $\Lambda_S$ from the conformal map $\xi(s)$ which sets the scale from where on the $s$-channel physics is integrated out. Explicitly, it is given by
\begin{eqnarray} \label{def-conformal}
\xi(s)=\frac{a\,(\Lambda^2_S-s)^2-1}{(a-2\,b)(\Lambda^2_S-s)^2+1}\,,\\
a = \frac{1}{(\Lambda^2_S-\mu_E^2)^2},\quad b = \frac{1}{(\Lambda^2_S-\Lambda^2_0)^2}\,,\nonumber
\end{eqnarray}
where the parameter $\Lambda_0$ is defined unambiguously such that the mapping domain of the conformal map touches the closest left-hand branch point. The expansion point $\mu_E$ identified with the s-channel thresholds. The parameter $\Lambda_S$ brings the main uncertainty in the prediction of \cite{Danilkin:2011fz, Danilkin:2012ap}. The finite value of $\Lambda_S$ indicates the energy above which other channels become important. We allow for a conservative variation of $\Lambda_S$ from $1.4$ GeV to $1.8$ GeV with the central value $\Lambda_S \simeq m_\rho+m_\omega=1.6$ GeV  determined by the point where the channel $\pi\eta \to \rho\,\omega$ opens up.

To identify correctly the mass and the width of the $a_0(980)$ resonance we search for poles in the complex $s$-plane. In the two-channel case there are four Riemann sheets, which correspond to different signs of the imaginary parts of the center of mass momenta \cite{Badalian:1981xj}. In the neighborhood of the pole, the $t$-matrix elements can be written as 
\begin{eqnarray}\label{t-pole}
t_{1,ij}^{0, \text{sheet}}(s)&\simeq& \frac{c_i\,c_j}{s^{\text{sheet}}_{a_0}-s}
\end{eqnarray}
where $s_{a_0}=(M_{a_0}\pm \frac{i}{2}\,\Gamma_{a_0})^2$. In Eq.(\ref{t-pole}) $i$ and $j$ are the coupled-channel indices and the couplings $c_{i}$, $c_{j}$ indicate the strength of coupling of the resonance to the each channel and may be related to partial-decay widths.

\begin{figure}[t]
\includegraphics[width =0.40\textwidth]{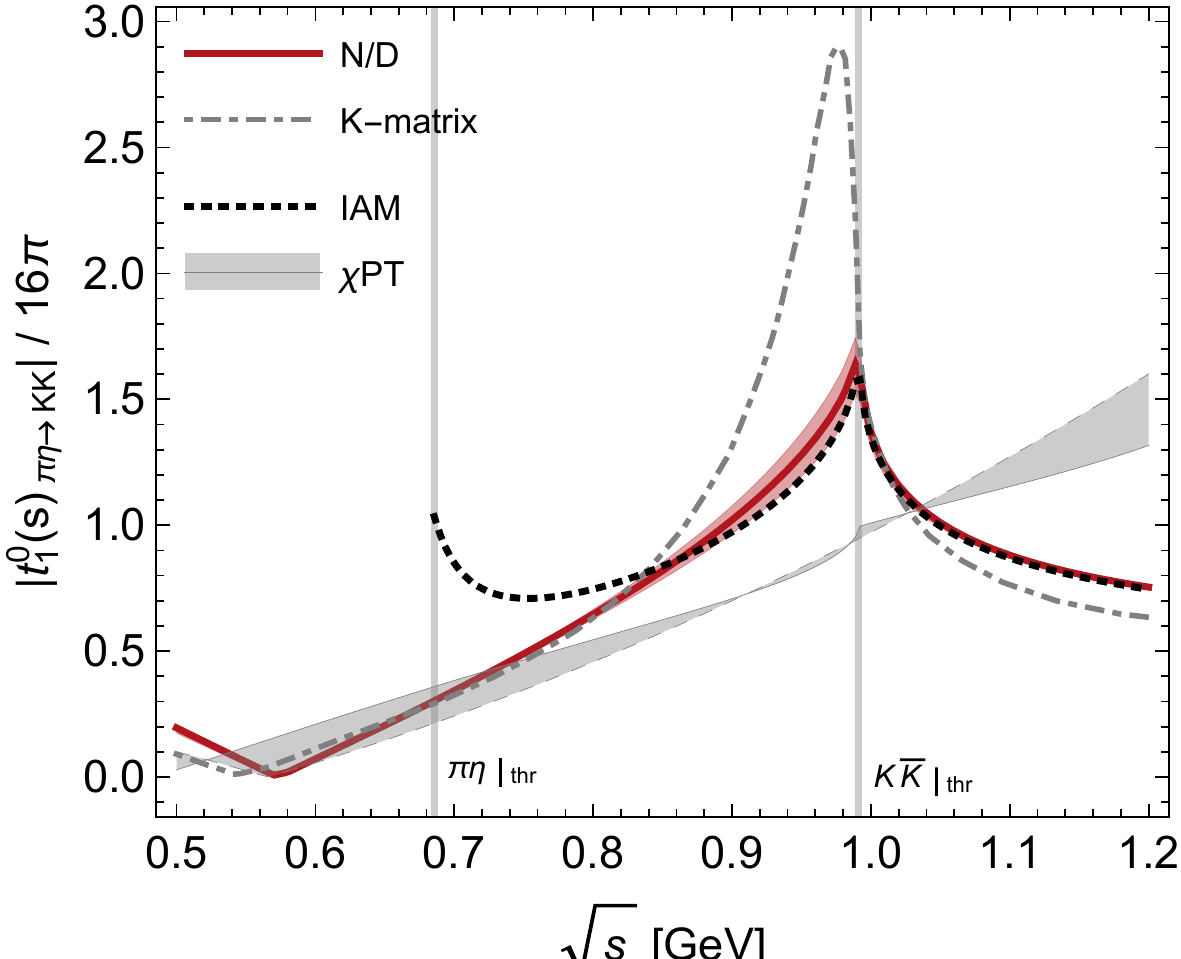}
\caption{The modulus of $t_{1}^0(s)_{\pi\eta \to K\bar{K}}$ from the dispersive approach (N/D) \cite{Danilkin:2011fz} compared to the K-matrix \cite{Albaladejo:2017hhj}, inverse amplitude method (IAM) \cite{GomezNicola:2001as} and $\chi$PT \cite{Gasser:1984gg,GomezNicola:2001as} analyses. For the latter we show the spread between the LO and NLO results (with the NLO low-energy constants taken from \cite{GomezNicola:2001as}).}
\label{fig:AbsT12}
\end{figure}

Performing the analytical continuation to the complex plane \cite{Gribov:1962fx}, we find a pole on the forth (IV) Riemann sheet (Sign(Im$\beta_{\pi\eta}$), Sign(Im$\beta_{K\bar{K}}$))=$(+,-)$)
\begin{eqnarray}\label{poles}
\sqrt{s^{\text{IV}}_{a_0}}&=&\left(1.12^{-0.07}_{+0.02}\right) -\frac{i}{2}\left(0.28^{+0.08}_{-0.13}\right)\,\text{GeV}
\end{eqnarray}
where the upper and lower error bars correspond to $\Lambda_S=1.4$ GeV and $\Lambda_S=1.8$ GeV, respectively. The residue of this pole leads to the ratio $|c_{K\bar{K}}/c_{\pi\eta}|=0.98^{-0.07}_{+0.20}$ which indicates a strong coupling of the $a_0(980)$ resonance to both the $\pi\eta$ and $K\bar{K}$ channels, as expected. We note that though the location of the pole is not too far away from the $K\bar{K}$ threshold, its precise determination may require taking into account higher order effects in the left-hand cuts of the dispersive approach \cite{Danilkin:2011fz}, or fitting the corresponding conformal expansion coefficients $C_k$ directly to future experimental data.

A number of theoretical efforts were devoted to understanding the properties of the $a_0(980)$ resonance \cite{Oller:1998hw, Oller:1998zr, GomezNicola:2001as, Guo:2011pa,Baru:2004xg, Albaladejo:2015aca}. Recently, the first analysis of the $\pi\eta$ scattering was performed by lattice QCD in Refs. \cite{Dudek:2016cru,Wilson:2016rid}. The finite volume spectra were fitted with a range of different K-matrix parametrizations and then analytically continued to complex energies. As a result, the pole on the IV sheet was found very close to the $K\bar{K}$ threshold. The analysis was conducted with the light quark masses corresponding to $m_\pi=391$ MeV. The extrapolation to the physical pion mass was performed recently within unitarized $\chi$PT \cite{Guo:2016zep}, which confirmed a pole on the fourth Riemann sheet. In all these cases, $a_0(980)$ shows up as a sharp (cusplike) peak exactly at the two kaon threshold. This behavior of the hadronic cross section is somewhat different from the recent K-matrix analysis \cite{Albaladejo:2015aca} which takes as an input the pole on the second Riemann sheet from \cite{Ambrosino:2009py, Isidori:2006we}. In Fig. \ref{fig:AbsT12} we compare the absolute value of the off-diagonal ($\pi\eta\to K\bar{K}$) scattering amplitude resulting from different approaches. This quantity has a particular importance for the $\gamma\gamma \to \pi \eta$ process since the rescattering through the intermediate $K^+K^-$ pair has a significant contribution to the cross section. One notices that our input from \cite{Danilkin:2011fz, Danilkin:2012ont} is consistent with $\chi$PT \cite{GomezNicola:2001as, Bernard:1991xb, Gasser:1984gg} and with the K-matrix approach \cite{Albaladejo:2015aca} at low energies, while in the resonance region it shows up as a prominent cusp similar to the result from the inverse amplitude method \cite{GomezNicola:2001as}.

\subsection{Left-hand cuts}
\label{sec:Left-hand_cuts}
\begin{figure}[t]
\includegraphics[width =0.3\textwidth]{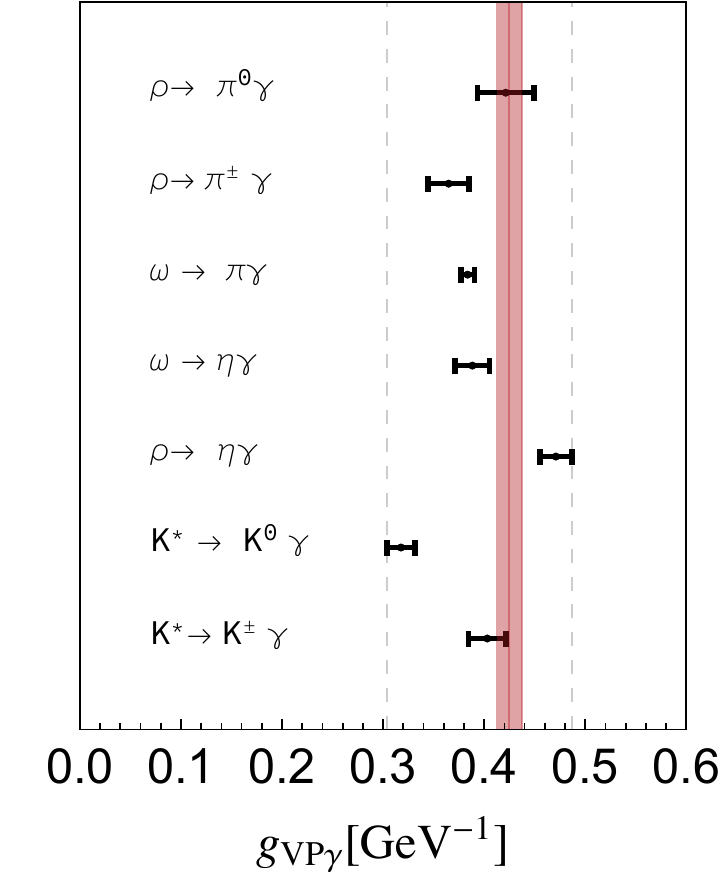}
\caption{Modulus of the radiative couplings scaled by SU(3) coefficients: \{$C_{\rho\to \pi^{0}\gamma}$, $C_{\rho\to \pi^{\pm}\gamma}$, $\frac{1}{3}C_{\omega\to \pi^0\gamma}$, $\sqrt{3}\,C_{\omega\to \eta\gamma}$, $\frac{1}{\sqrt{3}}C_{\rho\to \eta\gamma}$, $\frac{1}{2}C_{K^*\to K^0\gamma}$, $C_{K^*\to K^{\pm}\gamma}$\}. The vertical dashed lines indicate the spread of SU(3) breaking effects. The red band indicates the value for the universal (effective) coupling that we obtain from the description of the $\eta\to \pi^0\gamma\gamma$ decay.}
\label{fig:gVPgamma}
\end{figure}

With the known Omn\`es function, the input we need in Eq.(\ref{h_CC}) is the left-hand cuts. These are the so-called Born terms, which are nonzero only for the $\gamma\gamma \to K^+K^-$ and $\text{Disc}\,\{h^0_{1,++}(s'),\,\bar{k}^0_{1,++}(s')\}$ along the left-hand cut. While the former can be calculated from the scalar QED
\begin{eqnarray}\label{F12_Born}
F_1(s,t)&=&-\frac{4\,e^2 \left(t\,u-m_K^4+\left(t-m_K^2\right) \left(u-m_K^2\right)\right)}{s\left(t-m_K^2\right)\left(u-m_K^2\right)}\,,\nonumber\\
F_2(s,t)&=&-\frac{e^2}{\left(t-m_K^2\right) \left(u-m_K^2\right)}\,,
\end{eqnarray}
the latter we approximate by vector-meson exchange diagrams, which we expect to be the second important left-hand cuts contributions. We use the simplest Lagrangian which couples photon, vector (V) and pseudoscalar (P) meson fields
\begin{eqnarray}
{\cal  L}_{VP\gamma}=e\,C_V\,\epsilon^{\mu\nu\alpha\beta}\,F_{\mu\nu}\partial_\alpha\,V_\beta\,,
\end{eqnarray}
where $C_V$ are the radiative couplings, which we fix from the 2016 PDG values \cite{PDG-2016} for the partial widths of light vector mesons using
\begin{eqnarray}
\Gamma_{V\to P\gamma}=\alpha\,\frac{C_{V\to P\gamma}^2}{2}\frac{(M_V^2-m^2)^3}{3M_V^2}\,.
\end{eqnarray}
Here $\alpha \equiv e^2 / (4 \pi) \simeq 1/137$ is the fine structure constant. We present the absolute values  of $C_{V\to P\gamma}$ in Fig. \ref{fig:gVPgamma}, which we scaled by the corresponding SU(3) coefficients for easier comparison. For the universal coupling we estimate $g_{VP\gamma}=0.4(1)\,\text{GeV}^{-1}$, where the choice $g_{VP\gamma}=0.3\,\text{GeV}^{-1}$ reproduces the $K^*\to K^0 \gamma$ width and the value $g_{VP\gamma}=0.5\,\text{GeV}^{-1}$ reproduces the $\eta \gamma$ width of the $\rho$ meson. The relatively large spread in values indicates significant SU(3) breaking effects. 

The invariant amplitudes for the $t$- and $u$-channel vector-mesons exchanges read
\begin{eqnarray}\label{F12_V}
F_1(s,t)&=&-\sum_V 2\,e^2 C_{12}\left(\frac{t}{t-M_V^2}+\frac{u}{u-M_V^2}\right),\nonumber\\
F_2(s,t)&=&\sum_V \frac{e^2 C_{12}}{2}\left(\frac{1}{t-M_V^2}+\frac{1}{u-M_V^2}\right),
\end{eqnarray}
where $C_{12}\equiv C_{V\to P_1\gamma} C_{V\to P_2\gamma}$. Note that Eqs. (\ref{F12_Born}) and (\ref{F12_V}) preserve $t \leftrightarrow u$ symmetry due to Bose statistic of the photons. Using a simple relation for the helicity-0 amplitude $H_{++}=- s\,F_1/2$ we get the following s-wave amplitudes
\begin{eqnarray}
k^{0,Born}_{++}(s)&=&e^2\frac{4\,m_K^2}{s\,\beta_{K}(s)}\,\log\frac{1+\beta_K(s)}{1-\beta_K(s)}\,,\\
h^{0,Vexch}_{++}(s)&=&\sum_V 2\,e^2C_{12}\left(-\frac{M_V}{\beta_{\pi\eta}(s)}L_V(s)+s\right),\nonumber
\end{eqnarray}
where
\begin{eqnarray}
\beta_{K}(s)&=&\sqrt{1-\frac{4m_K^2}{s}},\quad L_V(s)=\log\frac{X_V(s)+1}{X_V(s)-1}\,,\nonumber\\
X_V(s)&=&\frac{2M_V^2-(m^2_\pi+m^2_\eta)+s}{s\,\beta_{\pi\eta}(s)}\,.\nonumber
\end{eqnarray}
The result for $k^{0, V exch}_{++}(s)$ can be obtained from $h^{0, V exch}_{++}(s)$ by replacing $m_{\pi,\eta}\to m_{K}$. From the logarithmic function one can see that the closest left-hand cut from the vector-meson exchange terms starts at
\begin{equation}
s_L=-\frac{(M_\rho^2-m_{\pi}^2)(M_{\rho}^2-m_{\eta}^2)}{M_{\rho}^2}\,.
\end{equation}
We also note that the p.w. vector-meson exchange amplitudes are not asymptotically bounded (they grow as $\sim s$). This is a consequence of the Lagrangian-based approach for the treatment of the left-hand cuts. There are several ways to overcome this problem. The usual way would be to introduce subtraction parameters that would suppress the high-energy behavior \cite{Moussallam:2013una}. A formal drawback, however, is that all these subtractions need to be fixed from the data or matched to $\chi$PT results. In addition, a subtraction polynomial of sufficient order will lead to an unphysical high-energy behavior and therefore severely limit the energy range of validity. To overcome these issues one can impose Regge constraints, which however require high-energy data in order to fix the parametrization. Another way, proposed in \cite{Gasparyan:2010xz}, is to use the conformal mapping technique. In the considered dispersive approach \cite{GarciaMartin:2010cw}, we emphasize that only the imaginary parts along the left-hand cut of the p.w. amplitudes are needed, which are asymptotically bounded, $\text{Im}\, h^{0, V exch}_{++}(-\infty) \to \text{const}$. Therefore, we will not introduce any modifications of the left-hand cuts in the present work. We also note that since our Omn\`es functions are asymptotically bounded at high energies \cite{Danilkin:2011fz}, one subtraction in Eq.(\ref{h_CC}) is sufficient for the convergence.

\subsection{$a_2(1320)$ resonance}
\label{sec:a2}
We approximate the $a_{2}(1320)$ resonance by a Breit-Wigner form, similar to how it was done for the $f_{2}(1270)$ resonance in \cite{Drechsel:1999rf, Hoferichter:2011wk}. It implies
\begin{equation}
h_{1,+-}^{2}(s)=-\frac{e^2\,C_{a_2\to\pi\eta}\,C_{a_2\to\gamma\gamma}\,s^2\,\beta^2_{\pi\eta}(s)}{10\,\sqrt{6}\,(s-M_{a_2}^2+i\,M_{a_2}\,\Gamma_{a_2}(s))}\,,
\end{equation}
where $C_{a_2\to \pi\eta}$ and $C_{a_2\to \gamma\gamma}$ couplings can be fixed from the experimental decay widths
\begin{eqnarray}
\Gamma_{a_2\to\pi\eta}&=&\frac{\beta_{\pi\eta}^5(M_{a_2}^2)}{1920\,\pi}\,C_{a_2\to \pi\eta}^2\,M_{a_2}^3\,,\nonumber\\
\Gamma_{a_2\to\gamma\gamma}&=&\frac{\pi\alpha^2}{5}\,C_{a_2\to \gamma\gamma}^2\,M_{a_2}^3\,,
\end{eqnarray}
assuming that the $a_2(1320)$ resonance is predominantly produced in a state with helicity-2. Using the PDG \cite{PDG-2016} values for the partial decay widths $\Gamma^{\text{PDG}}_{a_2\to\pi\eta}=15.5(1.5)$ MeV and $\Gamma^{\text{PDG}}_{a_2\to\gamma\gamma}=1.0(1)$ keV, the resulting couplings are
\begin{eqnarray}
|C_{a_2\to \pi\eta}| &=& 10.8(5) \, \mathrm{GeV}^{-1}, \nonumber \\
|C_{a_2\to \gamma\gamma}| &=& 0.115(5) \, \mathrm{GeV}^{-1}. 
\end{eqnarray}
For the parametrization of the total width $\Gamma_{a_2}(s)$ we follow the Belle Collaboration where the $a_2(1320)$ decays into $\pi\rho$, $\pi\eta$, $\omega \pi\pi$ and the $K\bar{K}$ final states were explicitly accounted for \cite{Uehara:2009cf}\footnote{We removed Blatt-Weisskopf factors which only slightly change the cross section in the considered region but introduce additional unknown parameter.}.

\subsection{$\eta \to \pi^0 \gamma\gamma$ decay}
\label{sec:Decay}
Crossing symmetry implies that the invariant amplitudes $F_{1,2}(s,t)$ describe not only the scattering process $\gamma\gamma \to \pi^0 \eta$ but also the decay process $\eta \to \pi^0 \gamma\gamma$. The differential decay rate is given by \cite{PDG-2016}
\begin{equation}
\frac{d^2\Gamma}{ds\,dt}=\frac{1}{(2\pi)^3}\frac{1}{32\,m_{\eta}^3}\sum_{\lambda_1,\lambda_2}|H_{\lambda_1\lambda_2}|^2\,,
\end{equation}
where crossing implies the following relations to the decay invariants $s\to M_{\gamma\gamma}^2$ and $t\to M_{\gamma\pi}^2$. 

\begin{figure}[t]
\includegraphics[width =0.45\textwidth]{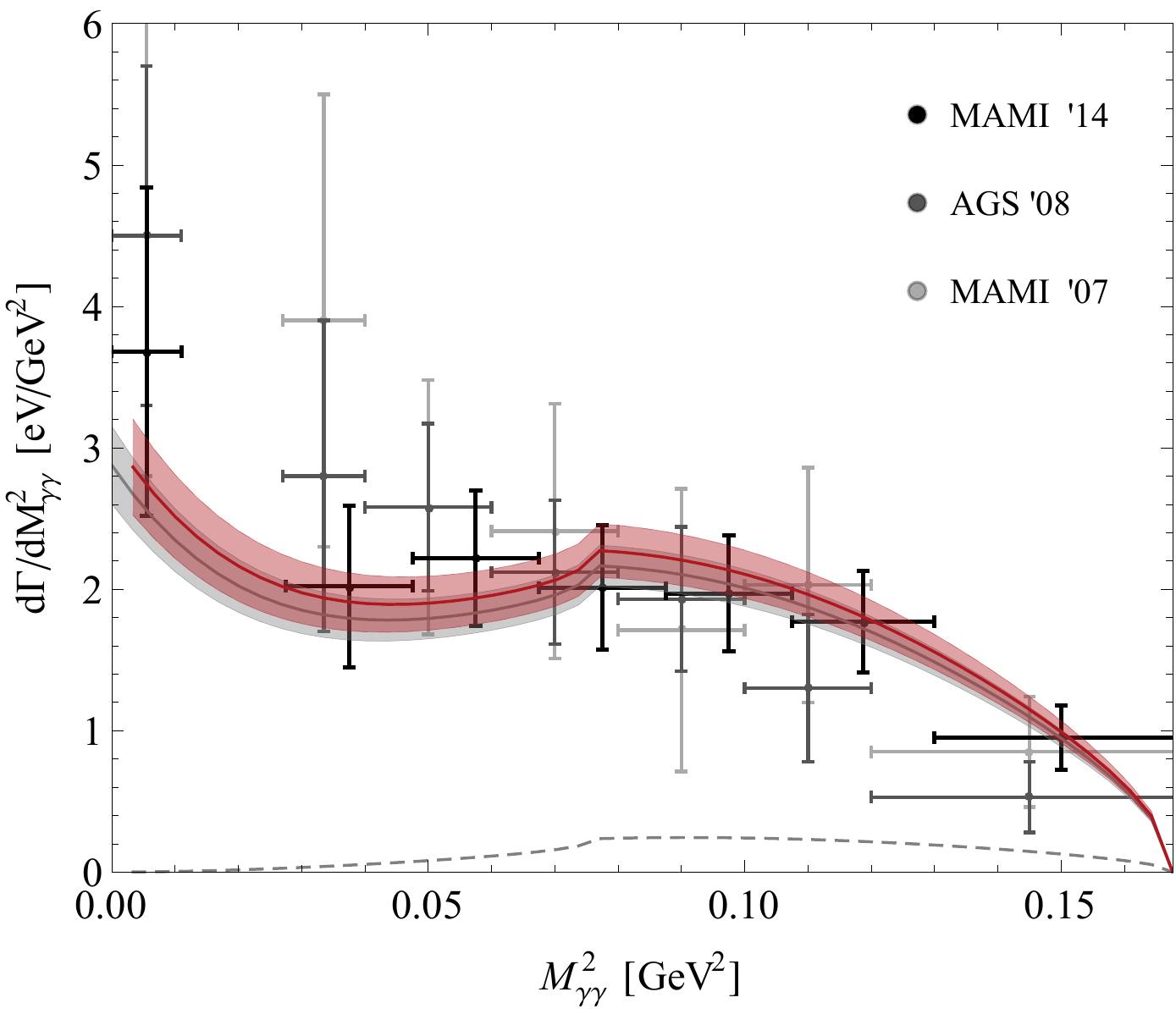}
\caption{The $d\Gamma/dM^2_{\gamma\gamma}$ distribution of the $\eta\to\pi^0\gamma\gamma$ decay. The gray band represents the result using physical radiative couplings $C_{V\to P \gamma}$, while the red band is a result of the fit with one universal (effective) coupling $g_{VP\gamma}$. The dashed curve indicates the $\chi$PT result at NLO. The data are taken from Refs.~\cite{Prakhov:2008zz,Prakhov:1900zz,Nefkens:2014zlt}.}
\label{fig:Decay}
\end{figure}

On the experimental side, the two-photon invariant mass distribution of this decay has been recently obtained by the A2 Collaboration at MAMI \cite{Nefkens:2014zlt}. This measurement has an improved statistical accuracy compared to previous measurements \cite{Prakhov:2008zz, Prakhov:1900zz}. Therefore, here we will only use the latest MAMI measurement and show the earlier data in Fig. \ref{fig:Decay} only for the reader's convenience.

Theoretically, this decay is traditionally used to test the higher order terms of $\chi$PT \cite{Gasser:1984gg}. The tree level amplitudes vanish both at leading (LO) and next-to-leading (NLO) orders. The first nonzero contributions come from either the pion or kaon loops \cite{Ametller:1991dp}. While the kaon loops are suppressed due to the large kaon mass, the contribution from the pion loops violates G-parity, and the decay amplitude is proportional to the small quantity $m_u-m_d$. The major contribution comes from the next-to-next-to-leading (NNLO) counterterms, which requires the knowledge of a set of low-energy constants. We saturate them using our vector $t$- and $u$- meson exchange terms. This would check the dynamical role of the vector mesons similar to \cite{Oset:2008hp,Danilkin:2012ua}.

\begin{figure*}[tb]
\begin{center}
\includegraphics[width =0.47\textwidth]{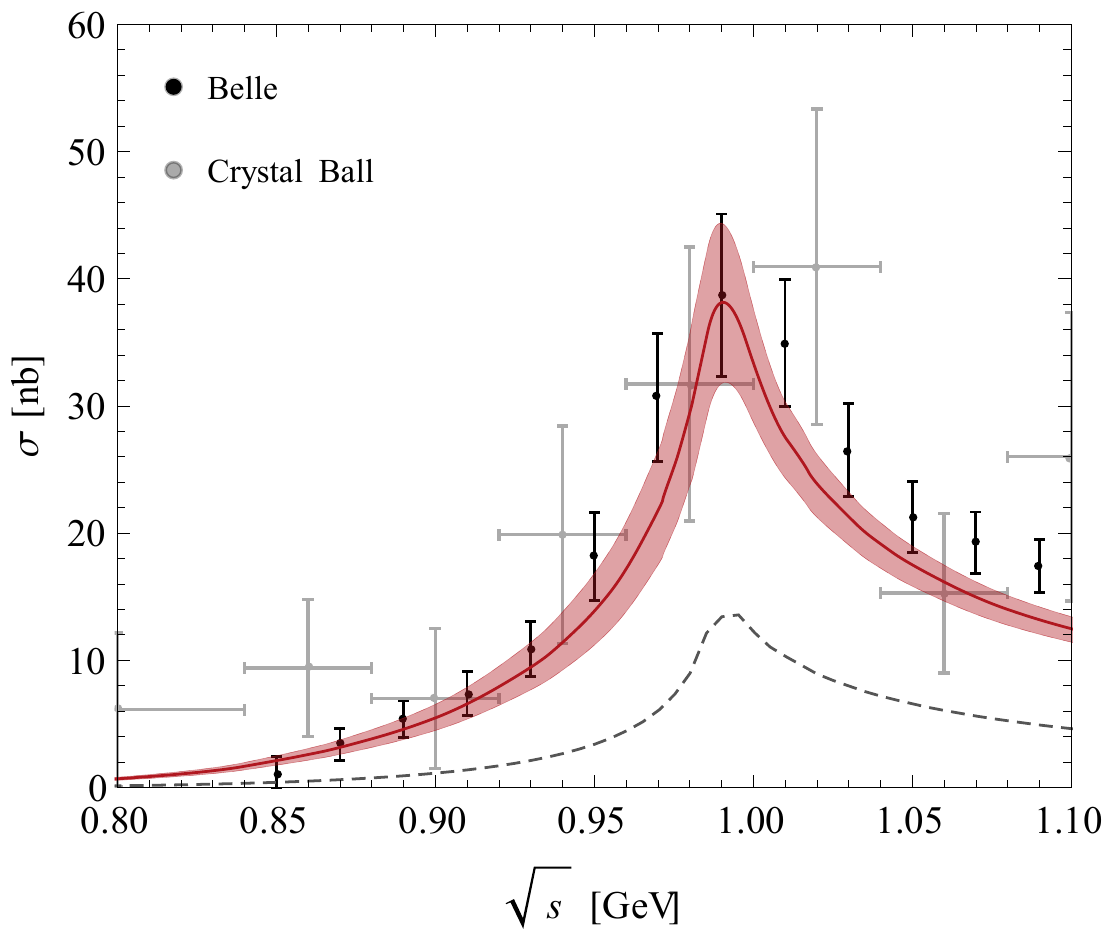}
\includegraphics[width =0.47\textwidth]{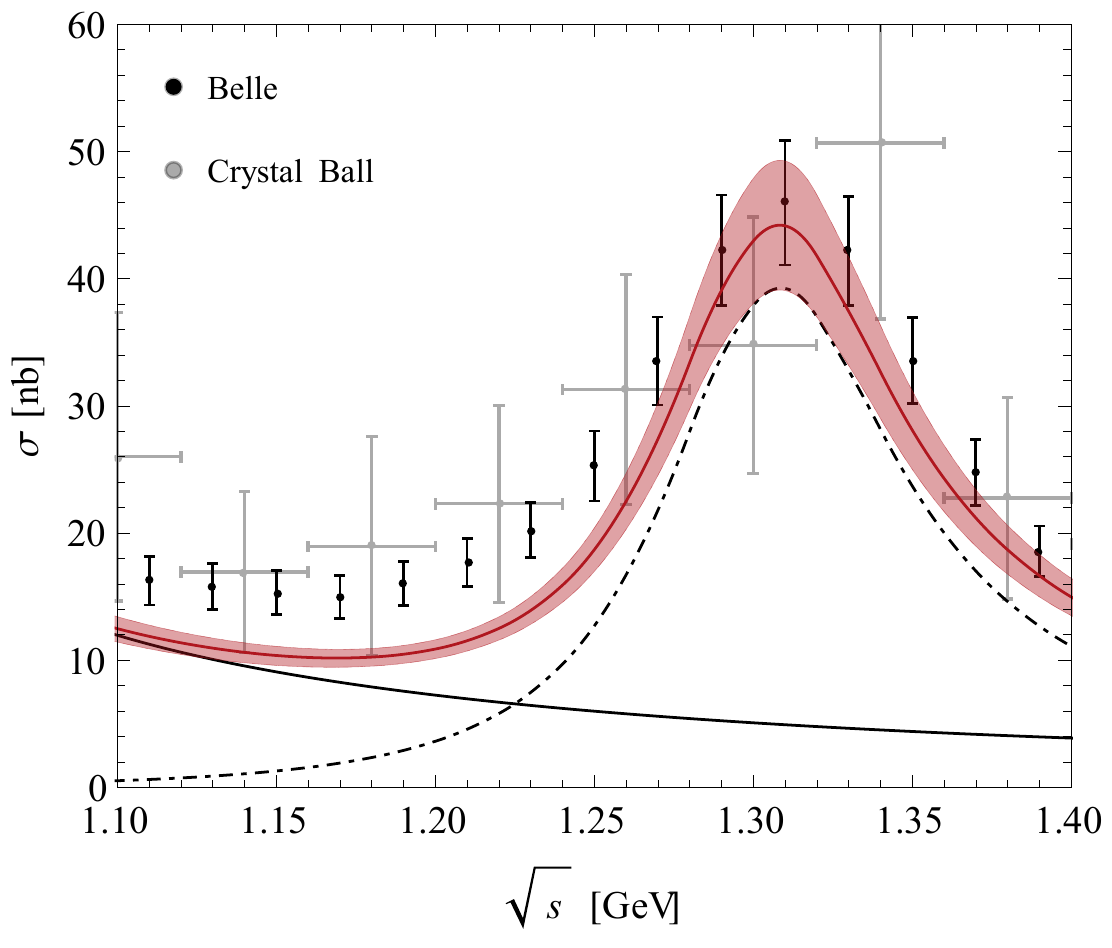}
\caption{A parameter-free postdiction for the $\gamma\gamma\to\pi^0\eta$ total cross section ($|\cos\theta|<0.8$). Left panel: Result of the dispersive representation, where we separately display the  rescattering contribution through the $K\bar{K}$ Born terms (dashed curve). Right panel: The sum of the $d$-wave Breit-Wigner parametrization (dashed-dotted curve) combined with the $s$-wave contribution (solid black line). A variation of the decay couplings $C_{\rho\to\pi^0\gamma,\eta\gamma,...}$ and $C_{a_2\to\pi\eta,\gamma\gamma}$ within their error bars and $1.4<\Lambda_S<1.8$ GeV is reflected by the shaded band. The data are taken from Refs.~\cite{Uehara:2009cf, Antreasyan:1985wx}.}
\label{fig:cross-section-postdiction}
\end{center}
\end{figure*}

While the vector-meson exchange contributions can be read off from Eq. (\ref{F12_V}), the $\chi$PT NLO loop contribution is taken from  \cite{Ametller:1991dp} and has the following form
\begin{equation}
F_{1}(s,t)=a^\pi+a^K,\quad F_{2}(s,t)=0\,,
\end{equation}
where
\begin{eqnarray}\label{api}
a^\pi&=&\frac{4\sqrt{2}\,\alpha}{3\sqrt{3}\,f^2}\Delta m_K^2\left(1+\frac{3(s-m_\pi^2)-m_\eta^2}{m_\eta^2-m_\pi^2}\right)I(s,m_\pi^2)\,,\nonumber\\
a^K&=&-\frac{2\sqrt{2}\,\alpha}{3\sqrt{3}\pi\,f^2}\left(3s-m_\eta^2-\frac{1}{3}m_\pi^2-\frac{8}{3}m_K^2\right)I(s,m_K^2)\,,\nonumber
\end{eqnarray}
with the loop function $I$ defined as
\begin{equation}
I(s,m^2)\equiv \int_{0}^1 dx \int _{0}^{1-x} dy \frac{xy}{m^2-s\,x\,y}\,.
\end{equation}
For the numerical estimates we use $f=92.3$ MeV and for the kaon mass difference in QCD we take $\Delta m_K^2=6.27(38)\times 10^{-3}\,\text{GeV}^2$ from \cite{Colangelo:2016jmc}. We find the kaon and pion loop contributions to the $\eta \to \pi^0 \gamma\gamma$ decay width as $\Gamma^{K\,loop}=0.010(0)$ eV and $\Gamma^{\pi\,loop}=0.003(0)$ eV, respectively. The latter was not included in the analysis of \cite{Oset:2008hp}, which in principle should be enlarged by the rescattering effects which are known to be strong for the $\eta \to 3\pi$ decay \cite{Guo:2016wsi,Guo:2015zqa,Colangelo:2016jmc,Schneider:2010hs,Albaladejo:2017hhj}. We leave this study for the future and take this contribution into account at the NLO level.

The individual contributions from the pion and kaon loops are relatively small compared to the PDG value~\cite{PDG-2016}: $\Gamma(\eta \to \pi^0 \gamma\gamma)=0.334(28)$. They can however interfere with vector-meson exchange terms. We find that the data favor a coherent interference. As can be seen in Fig. \ref{fig:Decay}, the latest MAMI data \cite{Nefkens:2014zlt} is described well, within the error bars, using physical radiative couplings, giving $\chi^2/d.o.f=1.92/7$ and $\Gamma(\eta \to \pi^0 \gamma\gamma)=0.291(22)\,\text{eV}$. In Fig. \ref{fig:Decay} we also show the results when the universal coupling of $V\to P \gamma$ is used. Since its error bar is pretty big, we fit its value to the two-photon invariant mass distribution. The fitted value of $g_{VP\gamma}$ will also account effectively for the contributions from the higher intermediate states. The fit slightly improves the description of the data with $\chi^2/d.o.f=1.60/(7-1)$ and $\Gamma(\eta \to \pi^0 \gamma\gamma)=0.303(29)\,\text{eV}$. The fitted value of the universal (effective) coupling is $g_{VP\gamma}=0.425(13)\,\text{GeV}^{-1}$ as shown in Fig. \ref{fig:gVPgamma}.

\section{Results}
\label{sec:Results}

\subsection{$\gamma\gamma \to \pi\eta, K\bar{K}$ cross sections}

To completely determine the helicity amplitudes for the 
$\gamma \gamma \to \pi \eta$ and $\gamma \gamma \to K \bar K$ processes,
we need to fix the subtraction constants in Eq.~(\ref{h_CC}). 
In this work, we match them to the field theory amplitudes, i.e.
\begin{eqnarray}
a&=&h_{1,++}^{0}(s_{th})\simeq h^{0, V exch}_{1,++}(s_{th}), \\
b&=&\bar{k}_{1,++}^{0}(s_{th})\simeq k^{0, V exch}_{1,++}(s_{th}).
\nonumber
\end{eqnarray}
In Fig. \ref{fig:cross-section-postdiction}, our parameter-free postdiction is confronted with the experimental data on the $\gamma\gamma \to \pi^0\eta$ cross section. The shaded areas in the figures indicate the uncertainties of the decay couplings $C_{\rho\to\pi^0\gamma,\eta\gamma,...}$ and $C_{a_2\to\pi\eta,\gamma\gamma}$ together with the error bar on  $\Lambda_S$. We note, that the proposed dispersive approach for the $a_0(980)$ resonance and a simple Breit-Wigner parametrization for the $a_2(1320)$ resonance yields already a reasonable agreement with the recent data from the Belle Collaboration \cite{Uehara:2009cf}. While the low-energy region is dominated by the $s$-wave partial wave, the region above 1.1 GeV is well described by the sum of the $a_2(1320)$ $d$-wave resonance and a tail from $a_0(980)$.

\begin{figure*}[tb]
\begin{center}
\includegraphics[width =0.47\textwidth]{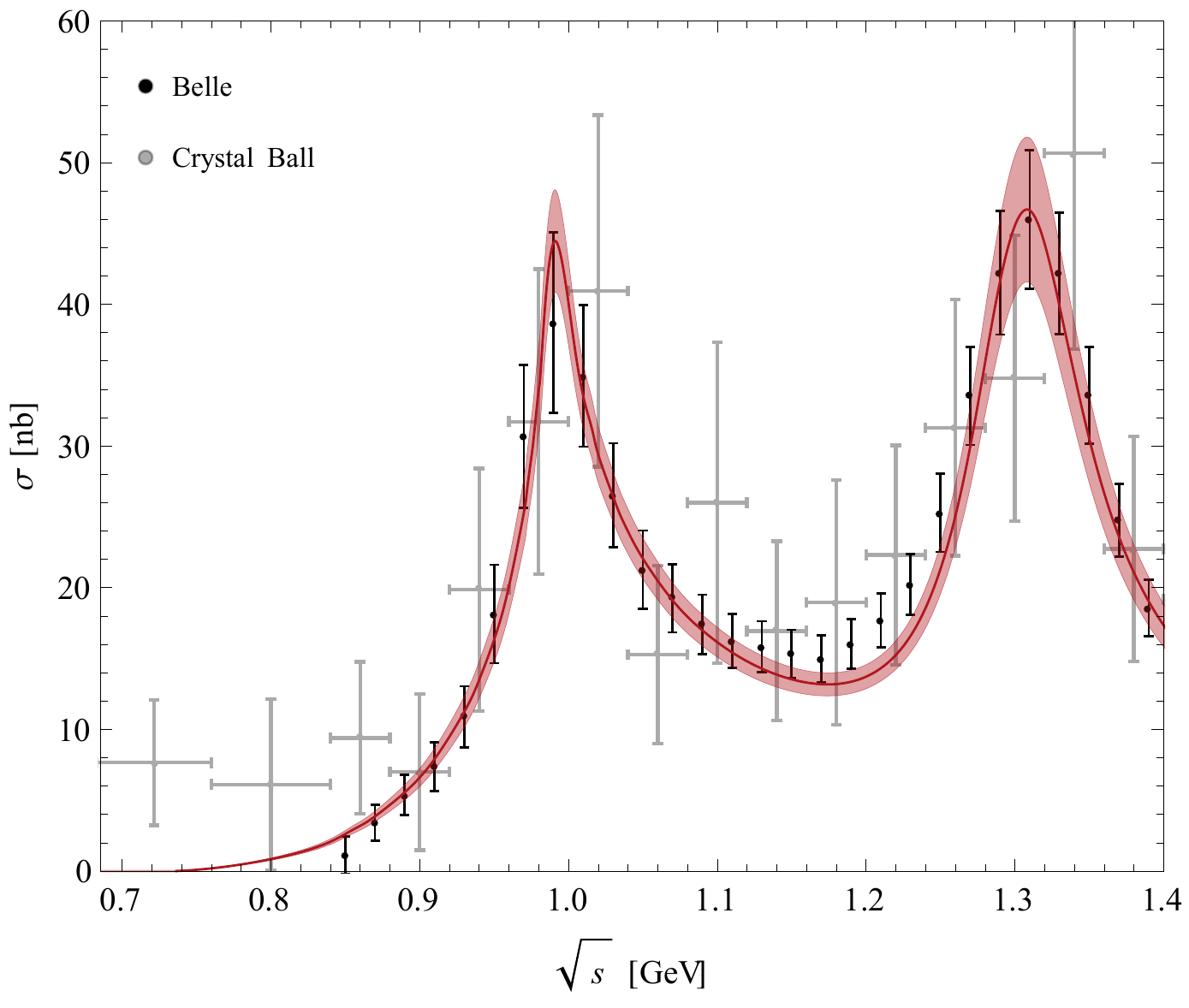}
\includegraphics[width =0.445\textwidth]{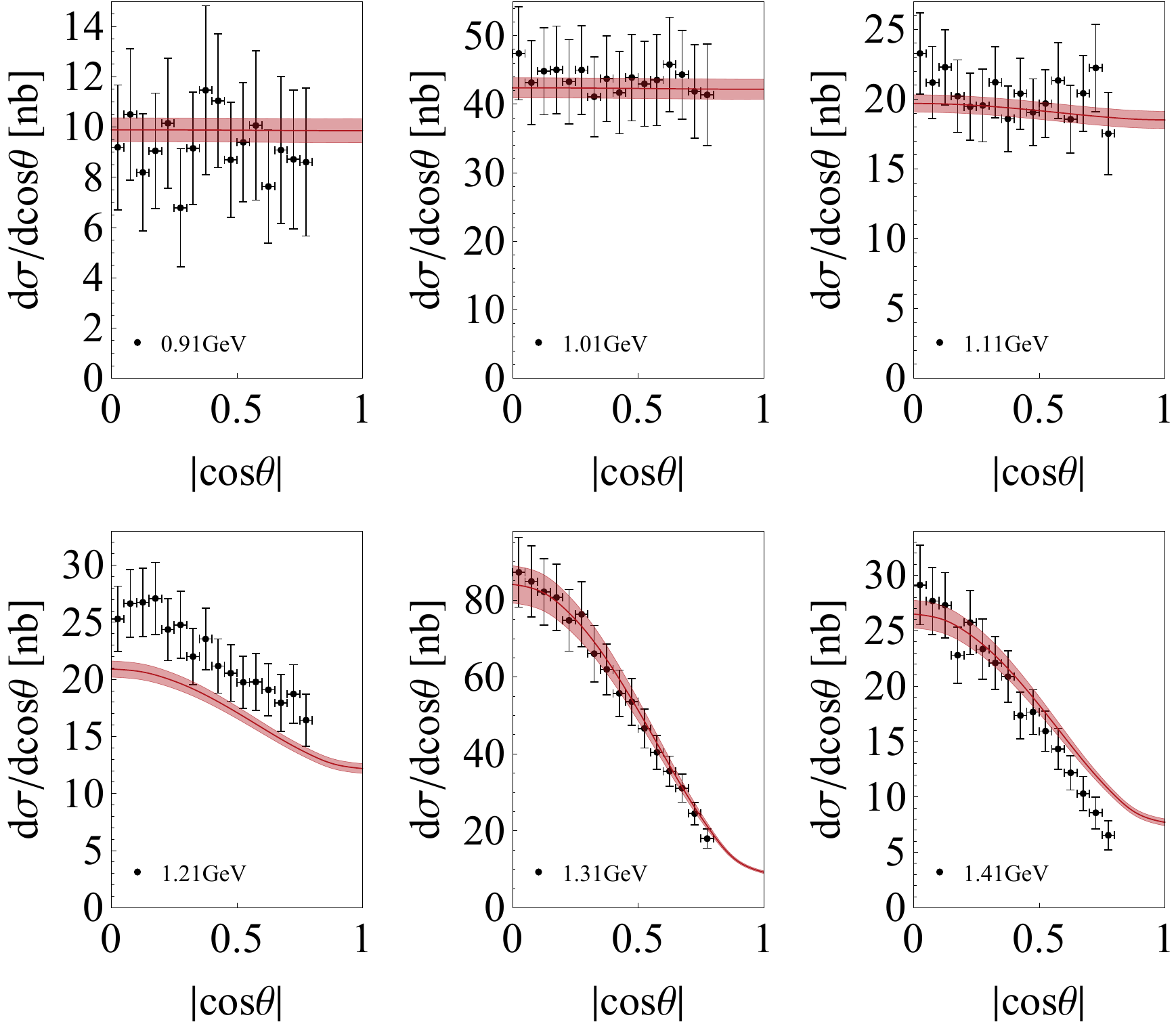}
\caption{Total (for $|\cos\theta|<0.8$) and differential cross sections for $\gamma\gamma\to \pi^0\eta$ using the universal (effective) coupling $g_{VP\gamma}$, 
and the fitted value of $\Lambda_S$, as explained in the text. The data are taken from Refs.~\cite{Uehara:2009cf, Antreasyan:1985wx}.}
\label{fig:cross-section-fit}
\end{center}
\end{figure*}

We further scrutinize the uncertainties of our approximation scheme. For this purpose, we firstly use the universal (effective) coupling  $g_{VP\gamma}$, which we constrained from the crossed process $\eta \to \pi^0 \gamma\gamma$. Secondly, one can use the existing cross section data to narrow down the uncertainty from the hadronic final state interaction, namely $\Lambda_S$. The fit to the Belle Collaboration data in the region $\sqrt{s}<1.1$ GeV leads to $\Lambda_S = 1.46(6)$ GeV and $\chi^2/d.o.f=0.34$. As a result, we obtain the description of the angular distributions and cross sections as shown in Fig.~\ref{fig:cross-section-fit}. We see that our results are in very good agreement with the data, except for slight 
disagreement in the differential cross section below and above the $a_2(1320)$ position. It can be improved most easily by taking $M_{a_2}=1.307$ GeV from the recent JPAC/COMPASS analysis \cite{Jackura:2017amb} rather than using the PDG average $M_{a_2}=1.318$ GeV \cite{PDG-2016}.

\begin{figure}[t]
\includegraphics[width =0.45\textwidth]{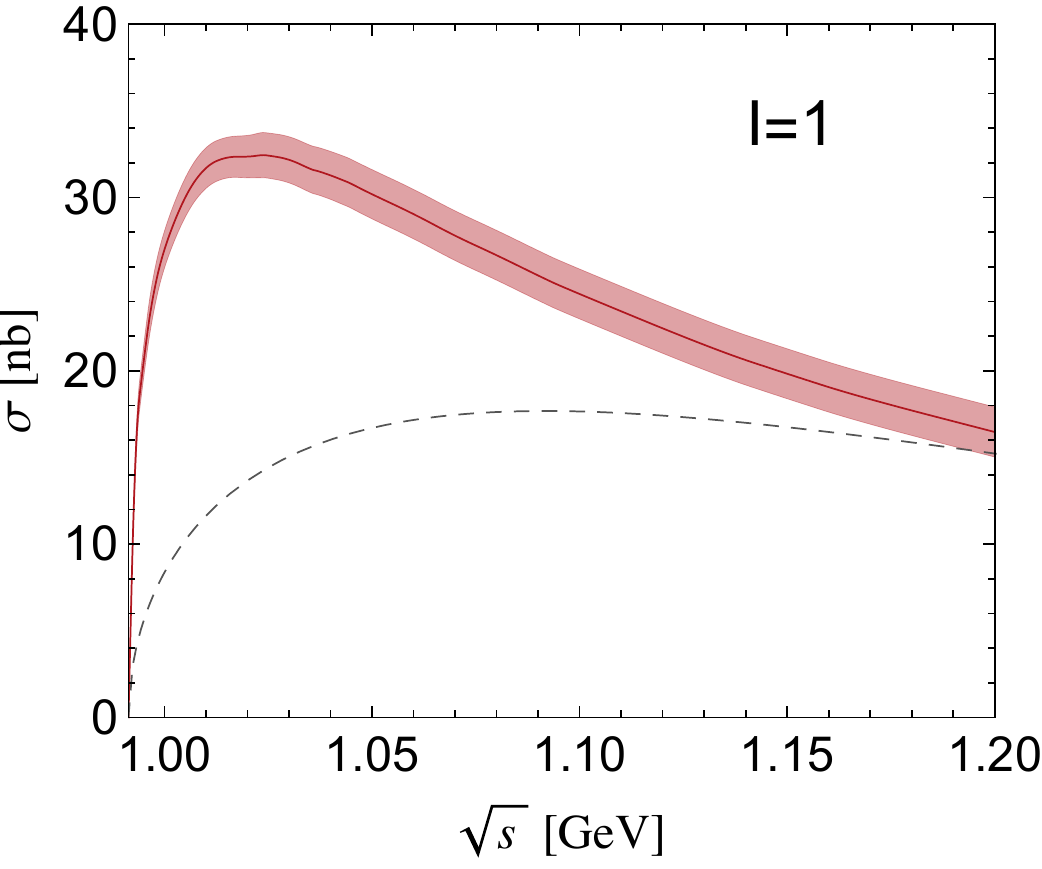}
\caption{Our prediction for the $\gamma\gamma\to K\bar{K}$ isovector total cross section. The dashed curve is the s-wave Born contribution.}
\label{fig:KK}
\end{figure}

In the coupled-channel treatment of Eq. (\ref{h_CC}), we have simultaneously calculated the isovector s-wave $\gamma\gamma \to K\bar{K}$ amplitude. This allows us to make a prediction for the corresponding $\gamma\gamma \to K\bar{K}$ cross section near the $K\bar{K}$ threshold. In Fig.\ref{fig:KK} we show
\begin{equation}\label{sigmaKK}
\sigma(s)=\frac{\beta_{K\bar{K}}(s)}{32\pi s} |k_{1,++}^0(s)|^2\,,
\end{equation}
compared to the pure Born result (i.e., when $k_{1,++}^0(s)$ is replaced by $k_{1,++}^{0, Born}(s)$). In both cases, we integrated the differential cross section over the whole angular range and neglected higher partial wave contributions. For the total result, we observe the cross section peaks close to the threshold indicating the presence of the $a_0(980)$ resonance.

Note that the isospin decomposition (\ref{IsospinK}) implies that the Born $I=0$ and $I=1$ $\gamma\gamma \to K\bar{K}$ amplitudes are the same. Therefore, the Born $\gamma\gamma \to K^+K^-$ cross section will be twice as large as the dashed curve shown in Fig. \ref{fig:KK}. On the other hand, the lower bound of the $\gamma\gamma \to K^+K^-$ cross section is half of (\ref{sigmaKK}) when neglecting the isoscalar contribution. The analysis of the latter is the subject of a separate paper. Based on previous result \cite{Danilkin:2012ap}, we expect that the isoscalar contribution will be suppressed. This is similar to the behavior in \cite{Oller:1998hw, Achasov:2009ee, Achasov:2012sc} where the drastic suppression of the Born term contribution was observed in the $\gamma\gamma \to K^+K^-$ channel due to final state interactions.

\subsection{Two-photon coupling of $a_0(980)$}

In order to extract the two-photon coupling of the $a_0(980)$ in our formalism, we can write in the neighborhood of the pole
\begin{eqnarray}
h_{1,++}^{0,\text{IV}}(s)&\simeq& \frac{c_{\gamma\gamma}\,c_{\pi\eta}}{s_{a_0}^{\text{IV}}-s}\,,
\nonumber
\end{eqnarray}
where $s_{a_0}^{\text{IV}}$ was obtained in Section \ref{sec:Hadronic_input}. The analytical continuation in the complex $s$-plane can be performed using the unitarity relation (similar to how it was done for the case of the second (II) Riemann sheet in \cite{Oller:2007sh}),
\begin{eqnarray}\label{hIV_unitarity}
h^{\text{IV}}(s)=h^{\text{I}}(s)+2\,i\,\rho_{K\bar{K}}(s)\,k^{\text{I}}(s)\,t_{K\bar{K}\to \pi\eta}^{\text{IV}}(s)\,,
\end{eqnarray}
where we suppressed isospin, spin and helicity indices for simplicity. In Eq.(\ref{hIV_unitarity}) the phase space factor proportional to the c.m. momenta $\rho_{K\bar{K}}(s)=\beta_{K\bar{K}}(s)/16\pi$ must be  analytically continued as well. Using Eq.(\ref{t-pole}) for the $K\bar{K}\to \pi\eta$ scattering amplitude on the IV Riemann sheet, one can express the two-photon coupling $c_{\gamma\gamma}$ through the hadronic $c_{K\bar{K}}$ coupling and the $\gamma\gamma\to K\bar{K}$ fusion amplitude, calculated at the resonance position on the first Riemann sheet:
\begin{eqnarray}
\left(\frac{c_{\gamma\gamma}}{c_{K\bar{K}}}\right)^2=-(2\rho_{K\bar{K}}(s_{a_0}^{\text{IV}}))^2\,(k^{\text{I}}(s_{a_0}^{\text{IV}}))^2\,.
\end{eqnarray}
In the narrow-width approximation, the radiative width is determined as\footnote{As pointed out in \cite{Dai:2014lza,Dai:2014zta}, this definition works well only for the narrow states which are well separated from the threshold cuts. In other cases Eq.(\ref{Gamma_a0}) serves as an intuitive way of re-expressing $|c_{\gamma\gamma}|^2$.} \cite{PDG-2016, Moussallam:2011zg}
\begin{equation}\label{Gamma_a0}
\Gamma_{a_0\to\gamma\gamma}=\frac{|c_{\gamma\gamma}|^2}{16\pi M_{a_0}}=0.27(4)\,\text{keV}\,.
\end{equation}
The obtained two-photon decay width in principle can be compared with the PDG value $\Gamma_{a_0\to\gamma\gamma}{\cal{B}}(\pi^0\eta)=0.21^{+0.08}_{-0.04}$ keV \cite{PDG-2016}. However, we like to emphasize that in all two-photon experimental analyses so far, the $a_{0}(980)$ peak has been approximated using a simple Breit-Wigner parametrization without any coupling to the $K\bar{K}$ channel \cite{Uehara:2009cf,Antreasyan:1985wx,Oest:1990ki}.

\section{Conclusions}
\label{sec:Conclusions}

In this work, we have presented a theoretical study of the $\gamma\gamma \to \pi^0 \eta$ reaction from the threshold up to 1.4 GeV in the $\pi\eta$ invariant mass. On the one hand, we used a coupled-channel dispersive approach in order to properly describe the scalar $a_0(980)$ resonance, which has a dynamical $\{\pi\eta, K\bar{K}\}$ origin. On the other hand, the $a_{2}(1320)$ tensor resonance has been introduced explicitly using a Breit-Wigner parametrization. 

The dispersive approach requires the knowledge of the amplitude on the left-hand cut. Beyond the well-known Born contribution we used $t$- and $u$-channel vector-meson exchanges with couplings fixed from experimental radiative decays of the vector mesons. This allowed us to show a parameter-free postdiction for the $\gamma\gamma \to \pi^0 \eta$ total cross section, which turned out to be in reasonable agreement with the recent empirical data from the Belle Collaboration. We have also tested the proposed treatment of the left-hand cuts using the crossed process, the $\eta \to \pi^0 \gamma\gamma$ decay. We have shown that NLO chiral perturbation theory supplemented with vector-mesons exchange terms reproduces the experimental two-photon invariant mass distribution very well. Moreover, in order to account for the contributions from the higher intermediate states, we have fitted the universal (effective) $g_{VP\gamma}$ coupling directly to the data. Consequently, we were left with the uncertainty coming from the hadronic final state interactions. Using the accurate Belle Collaboration data on the cross section, we narrowed down that error bar as well.

In order to extract the two-photon coupling of the $a_0(980)$ resonance, we analytically continued the amplitude into the unphysical regions. We found the pole on the fourth Riemann sheet, which produces a strong cusplike behavior of the cross section exactly at $K\bar{K}$ threshold. At the pole position, we calculated the two-photon coupling, and extracted the corresponding two-photon radiative width as $\Gamma_{a_0\to \gamma\gamma}=0.27(4)$ keV.

The obtained results can be used as a necessary starting point for a further study where one of the initial photons has a finite virtuality. The latter serves as one of the inputs to constrain the hadronic piece of the light-by-light scattering contribution to the muon's $(g-2)_\mu$ \cite{Pauk:2014rfa, Colangelo:2017fiz, Colangelo:2017qdm}. Its measurement is part of an ongoing dedicated experimental program at BESIII.

\section*{Acknowledgements}
I. D. acknowledges useful discussions with D. Wilson and J. Dudek. This work was supported by the Deutsche Forschungsgemeinschaft (DFG) in part through the Collaborative Research Center [The Low-Energy Frontier of the Standard Model (SFB 1044)], and in part through the Cluster of Excellence [Precision Physics, Fundamental Interactions and Structure of Matter (PRISMA)]. This work was also supported partially through GUSTEHP.

\bibliographystyle{prsty}
\bibliography{ggtoeta_short_2}

\end{document}